# Molecular structure, vibrational, photophysical and nonlinear optical properties of L-threoninium picrate: A first-principles study

*S. AlFaify[a]\*, Mohd. Shkir[a]\*, M. Arora[b], Ahmad Irfan[c,d], H. Algarni[a], Haider Abbas[e], Shabbir Muhammad[a], Abdullah G. Al-Sehemi[c,d]*

[a]*Department of Physics, Faculty of Science, King Khalid University, P.O. Box. 9004 Abha 61413, Saudi Arabia*
[b]*CSIR- National Physical Laboratory, Dr. K.S. Krishnan Road, New Delhi 110012, India*
[c]*Department of Chemistry, Faculty of Science, King Khalid University, Abha 61413, P.O. Box 9004, Saudi Arabia*
[d]*Research Center for Advanced Materials Science, King Khalid University, Abha 61413, P.O. Box 9004, Saudi Arabia*
[e]*Department of Physics, Manav Rachna College of Engineering, Faridabad, Haryana 121001, India*

*Corresponding author[\*]*
*Dr. Mohd. Shkir*
E-mail: shkirphysics@gmail.com, shkirphysics@kku.edu.sa,
Contact: +966 530683673, Fax: + 966 72418319

**Abstract**

In this work, different computational methods such as HF, B3LYP, range separated functionals (CAM-B3LYP and LC-BLYP) with 6-31G* basis set were applied to investigate the electronic, spectroscopic and nonlinear optical properties of L-threoninium picrate (LTHP) molecule for the first time. The calculated values of IR and Raman vibrational frequencies were found to be in a good agreement with the experimental results. Time dependent density functional theory has been applied to calculate the electronic and photophysical properties such as excitation energy, dipole moment and frontier molecular orbital (FMO) energies of LTHP molecule. The excitation energy value calculated by CAM-B3LYP is found to be at 351 nm which is in close agreement with the experimental values. The total/partial DOS (T/PDOS) was determined using GGA/BLYP. The total dipole moment ($\mu_{tot}$), static total and anisotropy of polarizability ($\alpha_{tot}$, $\Delta\alpha$) and static first hyperpolarizability ($\beta_0$, $\beta_{tot}$) values were calculated and compared with the reference compound. The $\mu_{tot}$ and $\beta_{tot}$ are found to be 3 and 51 time higher than urea molecule respectively. The FMOs, molecular electrostatic potential (MEP), global reactivity descriptors were also calculated and discussed. All these results suggest that the L-threoninium picrate would be a good candidate for optoelectronic device applications.

## 1. Introduction

In recent years, the advancement in organic nonlinear optical (NLO) materials has increased enormously because of their impending applications in optoelectronic, electro-optic, photonic, semiconductors, ferroelectric and superconductors devices [1-4]. On the supramolecular level to demonstrate high second harmonic generation (SHG) output, the material should have the structure of noncentrosymmetry, very strong intermolecular interactions and good phase-matching capability. On other hand the molecular scale, the amount of charge transfer is assumed to dominate the SHG output [5-7]. The materials which can transform the various parameters of optical beams such as frequency, phase, amplitude and polarization etc. are very important for NLO activity. The variety of materials including organic, inorganic, semiorganic, organometallic and polymeric have been studied for NLO applications. The organic materials receive the maximum attention because they are found to exhibit very high NLO response over a wide frequency range due to the presence of active $\pi$ bonds. In addition to the above properties, these materials have large structural diversity, inherent synthetic flexibility, large optical damage threshold for laser power, low frequency dispersion, ultrafast response, photo stability, large first hyperpolarizability ($\beta$) and permit to modify the chemical structures and properties to achieve the desired NLO properties [8-10].

Picric acid acts as an electron acceptor and having very good tendency to synthesize the stable picrate complexes with a variety of organic compounds through strong hydrogen bonding or $\pi$ interaction such as glycine, L-asparagine, 2-aminopyridine, 8-Hydroxyquinoline, L-proline and L-leucine etc. Their single crystals were also grown and studied previously [11-15]. These picrate compounds are found to show very large NLO response as well. Recently, a picrate compound named L-threoninium picrate (LTHP) has been synthesized by Natrajan et al. [16] and its crystal growth, structural, infrared, optical, and second harmonic generation efficiency (SHG) properties [SHG efficiency of LTHP was found to be 43 times that of standard potassium dihydrogen phosphate (KDP) crystals have been reported experimentally [16].

The quantum chemical approaches such as Hartree-Fock (HF) and density functional theory (DFT) which are cost effective general methods for analyzing almost all important properties very accurately [17, 18]. The main advantages of HF and DFT are to determine and reproduce the reasonable and accurately molecular geometries and vibrational frequencies. Moreover, the range separated functionals such as CAM-B3LYP and LC-B3LYP are efficient to calculate the electronic and NLO properties which are much superior to the conventional methods [19-25]. The current available literature on the titled compound shows that there is no report on its theoretical studies so far. Due to excellent NLO response and other properties in the titled compound, its theoretical investigations seem to be necessary and justified to get deep insight of it for future device applications. Hence, in the current work our main aim is to report the various needful properties such as molecular geometry, electronic, infrared, vibrational, NLO and optical properties of recently developed compound L-threoninium picrate (LTHP) have been investigated using various computational methods. The calculated results were explained and compared with the experimental wherever available.

## 2. Computational details

The optimization of ground state molecular geometry [26-28] and vibrational frequency calculations for LTHP molecule have been performed at HF, B3LYP and CAM-B3LYP with 6-31G$^*$ basis set. Further, the electro-optical properties such as excitation energies, dipole moment and NLO properties of LTHP molecule have been calculated. HF and B3LYP methods were chosen because they are better methods for computation of molecular structures, IR spectra and Raman spectra. The time dependent density functional theory (TDDFT) using B3LYP and range separated functionals (CAM-B3LYP and LC-BLYP) has been applied to calculate the optical properties [20-22]. All the theoretical calculations on LTHP molecule have been performed with Gaussian G09 package software with the default convergence criteria, without any constraint on the geometry. To analyze all stationary points as minima for Infra-red (IR) and vibrational (Raman) frequency calculations at DFT levels the optimized structural parameters were used. By taking the second derivative of energy the infrared and Raman spectra of LTHP were attain and computed analytically [29, 30].

Various other important parameters such as FMOs, dipole moment, polarizability, first hyperpolarizability and optical properties of LTHP were calculated. Finite Field (FF) method was applied to calculate the static first hyperpolarizability ($\beta_{tot}$) and its tensor components for titled compound. The FF method was mostly applied to study the NLO properties because it can be used in concert with the electronic structure method to work out $\beta$ values. Recently, the calculated $\beta_{tot}$ by this method is found to be legitimate with experimental structure property relationship. In this method, a molecule is subjected to a static electric field (*F*) then the energy (*E*) is expressed by the following equation:

$$E = E^{(0)} - \mu_1 F_1 - \frac{1}{2}\alpha_{ij}F_iF_j - \frac{1}{6}\beta_{ijk}F_iF_jF_k - \frac{1}{24}\gamma_{ijkl}F_iF_jF_kF_l - \cdots$$

where all the symbols are having their usual meanings and by differentiating this equation with respect to *F* the values of μ, α, β, and γ can be acquire.

## 3. Results and discussion

### 3.1. Molecular geometry

The optimized ground state molecular geometry of LTHP molecule was obtained by HF, B3LYP, and range separated functionals (CAM-B3LYP and LC-BLYP) with 6-31G* basis set. The geometry optimized by B3LYP and CAM-B3LYP using 6-31G* basis set along with numbering scheme has been figured out in Fig. 1 (a) and (b). The comparative important geometrical parameters such as bond lengths and bond angles are presented in Table 1 and found in close agreement with experimentally calculated values [16]. The hydrogen bond in LTHP molecule is shown by dotted lines [Fig. 1]. Two strong hydrogen bonds were observed between picric acid and L-threonine molecules. The one hydrogen bond between H(11) and O(29) atom i.e., N(10)–H(11)---O(29) with bond length 1.561 Å and the second hydrogen bond between O(35) and H(36) i.e., O(15)–H(36)---O(35) with bond length 1.764Å. While these bond lengths calculated by CAM-B3LYP [Fig. 1(b)] were found to be 1.534 and 1.768 Å, respectively. These bond length values show that the hydrogen bond is very strong between L-threonine and picric acid. An intra-molecular hydrogen bond between the amine N atom and the nearby oxygen atom, viz., N10–H12---O16, N10–H13---O14 and C2–H3---O16 were also observed with bond lengths 2.058, 2.665 and 2.643 Å, while these bond lengths calculated by CAM-B3LYP [Fig 1(b)] are found to be 2.047, 2.597 and 2.632 Å, respectively. The calculated bond lengths are very close to the experimental values [16].

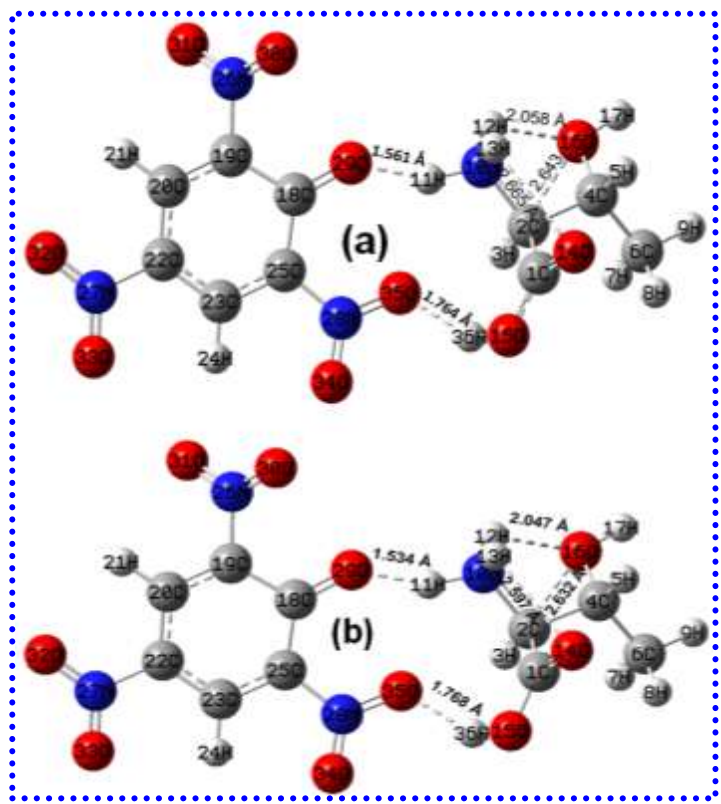

**Fig. 1**. Optimized ground state geometry of L-threonine picrate (LTHP) obtained with (a) B3LYP and (b) CAM-B3LYP using 6-31G$^*$ basis set.

**Table 1** The calculated bond lengths and bond angles of LTHP optimized molecule by DFT/B3LYP using 6-31G* level of theory

| Bond lengths | Cal. (Å) | Bond Angles | Cal. (°) | Bond lengths | Cal. (Å) | Bond Angles | Cal. (°) |
|---|---|---|---|---|---|---|---|
| L-threonine cation | | | | Picrate anion | | | |
| C1-O14 | 1.0196 | O14-C1-O15 | 123.322 | C18-O29 | 1.2525 | C19-C18-O29 | 121.093 |
| C1-O15 | 1.3126 | O14-C1-C2 | 119.037 | C18-C19 | 1.4598 | C25-C18-O29 | 126.404 |
| C1-C2 | 1.2140 | C1-C2-H3 | 112.946 | C19-N26 | 1.4680 | C19-N26-O30 | 118.164 |
| C2-H3 | 1.5420 | H3-C2-N10 | 106.083 | N26-O30 | 1.2277 | C19-N26-O31 | 117.005 |
| C2-C4 | 1.0888 | C2-N10-H11 | 114.615 | N26-O31 | 1.2336 | C19-C20-H21 | 120.367 |
| C4-H5 | 1.0917 | C2-N10-H12 | 107.237 | C19-C20 | 1.3715 | H21-C20-C22 | 120.203 |
| C4-C6 | 1.5076 | C2-N10-H13 | 110.411 | C20-H21 | 1.0827 | C22-N27-O32 | 117.456 |
| C6-H7 | 1.0313 | C2-C4-H5 | 107.937 | C20-C22 | 1.4047 | C22-N27-O33 | 117.786 |
| C6-H8 | 1.0229 | C2-C4-C6 | 111.943 | C22-N27 | 1.4573 | C22-C23-H24 | 120.172 |
| C6-H9 | 1.0625 | H5-C4-O16 | 110.410 | N27-O32 | 1.2326 | C22-C23-C25 | 119.889 |
| C4-O16 | 1.0955 | C4-C6-H7 | 110.727 | N27-O33 | 1.2325 | C24-C23-C25 | 119.980 |
| O16-H17 | 0.9694 | C4-C6-H8 | 111.036 | C22-C23 | 1.3803 | C23-C25-N28 | 115.720 |
| C2-N10 | 1.5004 | C5-C6-H9 | 110.124 | C23-H24 | 1.0820 | O25-N28-O34 | 120.246 |
| N10-H11 | 1.0822 | C2-C4-O16 | 104.129 | C23-C25 | 1.3972 | O25-N28-O35 | 118.405 |
| N10-H12 | 1.0295 | C4-O16-H17 | 108.870 | C25-N28 | 1.4305 | C18-C25-N28 | 121.240 |
| N10-H13 | 1.0258 | C4-C2-N10 | 108.630 | N28-O34 | 1.2290 | | |
| | | | | N28-O35 | 1.2600 | | |

The Mulliken charge in any molecule is directly related to their vibrational properties and quantifies how the electronic structure charges under atomic displacement. Therefore it is directly related to the chemical bonds present in the titled molecule. It affects many parameters of the molecule such as its dipole moment, polarizability, electronic structure and other properties of molecular system [31, 32]. The net atomic charges of LTHP molecule in ground state are tabulated in Table S1 and also shown in Fig. S1 (see supplementary data). While all the hydrogen atoms in the molecule have positive charges and very similar negative charges are noticed for the oxygen atoms. The carbon atom C(1) acquire maximum positive charge due to its bonding with two electronegative oxygen atoms O(14) and O(15) [33, 34]. It is important to mention here that N(10) atom has the maximum negative charge because may be it draws charge from its adjacent hydrogen atoms.

## 3.2. Vibrational analysis

*L*-Threoninium picrate ($C_{10}H_{12}N_4O_{10}$) crystallize in the monoclinic non-centrosymmetric space group *P*21 (*Z*=2) with unit cell dimensions a = 9.235(5) Å, b = 6.265(5) Å, c = 12.727(5) Å, β = 108.42(5)° and cell volume 698.6(7) Å$^3$ [16]. The unit cell comprises *L*-threoninium cations and picrate anions (Fig. 1). *L*-threoninium cation has carboxyl (-COOH), a protonated $NH_3^+$, methyl and alcoholic -OH groups while picrate anion has o- and p- substituted nitro (-$NO_2$), alcohol (-OH) groups and phenyl ring. The loss of H in the picrate anion is confirmed by the C–C distances near the phenolic group (C5–C6 and C5–C10). The picrate anion forms a strong asymmetric intermolecular; O–H...O and N–H...O hydrogen bonds with the threonine cation. Out of the three nitro groups, two (N2/O5/O6 and N4/O9/O10) present at ortho-positions w.r.t. C-OH group which are twisted from the plane of ring. While, C-N twisting vibration is not C-N bond length dependent [35]. The molecular structure proposed by Natarajan et al. of LTHP is presented in Fig. 1. Intermolecular N1–H1C...O1 hydrogen bonding interaction form a linear zigzag head-to-tail chain of amino acid. These linear chains propagates along the b-axis which are separated by picrate anions on either side through O2–H2...O4 inter-molecular hydrogen bonding interaction. An intra-molecular hydrogen bonds are formed between the amine N atom and the nearby oxygen atom, e.g. N1–H1B...O3 and within different zigzag structures of chain. The vibrational (IR and Raman) spectroscopy play a vital role in revealing the structural details of organic, inorganic and hybrid compounds from the appearance of different fundamental vibrational modes of basic skeletal and functional groups present in a material. In organic compounds, the conformational arrangement of molecular groups sometimes may lead to shifting of some peak positions to higher or lower frequencies with gain or loss in intensity. Theoretically, IR and Raman spectra of LTHP molecule are derived by B3LYP and HF approaches using 6-31G* basis set are presented Figs. 2 (a-c) and 3 (a-c), respectively. The observed IR and Raman peaks of LTHP functional groups and skeletal molecule are tentatively assigned on the basis of earlier reported available data on amino acids and picrate compounds [36, 37]. Due to the combination of electron correlation effects and basis set deficiencies, the calculated harmonic frequencies are found to be higher than the experimentally observed frequencies. So, for 6-31G* basis set of B3LYP and HF the scaling factor 0.9613 and 0.8929 were used, respectively. Table S2 (see supplementary data) shows all the calculated frequencies and their tentative assignments which are in well correlation with the reported values on such type of picrate compounds [16, 37, 38]. The O-H bond in phenol of picrate is broken down and the hydrogen is attracted by the N of the amino group to form $NH_3^+$ in L-Threonine. The removal hydrogen from O-H group, the double bond is created between C and O forms C=O.

**In 3400 – 3000 $cm^{-1}$ Region**

In amino acid, -OH group stretching vibrations are generally observed around 3500 $cm^{-1}$. Their peak positions are very sensitive to the environment and exhibit pronounced shift due to hydrogen bonded species. The asymmetric stretching vibrational mode of O–H group appear as medium to strong intensity broad band in IR spectrum. In these spectra, a strong O–H stretching mode band appeared at 3340 $cm^{-1}$ (B3LYP), 3348 (HF) in the IR spectra and at 3334 (B3LYP), 3348 (HF) in Raman spectra. This mode is merged with the –$NH_2$ asymmetric stretching mode and both appeared at the same position. In IR transmission spectra, N-H asymmetric and symmetric stretching modes of amine (-$NH_2$) group appear as strong and medium intensity peaks at 3340 $cm^{-1}$ and 3271 $cm^{-1}$ (B3LYP) and at 3348 $cm^{-1}$ and 3276 $cm^{-1}$ (HF), respectively. While in Raman spectra these modes are obtained as weak intensity peaks at 3334 $cm^{-1}$ and 3265 $cm^{-1}$ (B3LYP) and at 3348 $cm^{-1}$ and 3277 $cm^{-1}$ (HF). The asymmetric stretching vibration –$NH_3^+$ cation is expected in 3200-3100 cm$^{-1}$ region; in *L*THP crystal IR and Raman spectra at B3LYP is obtained as weak intensity peak at 3150 cm$^{-1}$ and 3135 cm$^{-1}$, respectively. C-H asymmetric and symmetric stretching vibration modes of aromatic benzene ring are observed for e$_{2u}$ symmetry vibrations at 3037 cm$^{-1}$, 3011 cm$^{-1}$ (B3LYP) and a$_{1g}$ non-degenerate symmetry vibration at 3089 cm$^{-1}$ (HF) in theoretical IR transmission spectra and for e$_{2g}$ symmetry vibration at 3031 cm$^{-1}$ (B3LYP), 3090 cm$^{-1}$ (HF) of a$_{1g}$ symmetry vibration in Raman spectra of this crystal [39, 40]. The methyl (-CH$_3$) group asymmetric stretching mode is observed at 2920 cm$^{-1}$ (B3LYP) and only symmetric stretching mode at 2837 cm$^{-1}$ (HF) in IR transmission spectra while in Raman spectra, asymmetric and symmetric stretching modes are obtained at 2945 cm$^{-1}$ and 2884 cm$^{-1}$ (B3LYP) and only asymmetric stretching mode at 2980 cm$^{-1}$.

**In 2000 - 1300 cm$^{-1}$ region**

In this region, the tentative vibrations of carbonyl (-COOH) group stretching modes, protonated amino (NH$_3$), amino (-NH$_3^+$) cation and methyl deformation modes, nitro (-NO$_2$) group stretching modes are observed. The carbonyl –C=O stretching mode is observed as strong peak at 1780 cm$^{-1}$ (B3LYP), 1810 cm$^{-1}$ (HF) infrared transmission spectrum and at 1777 cm$^{-1}$ (B3LYP), 1812 cm$^{-1}$ (HF) as weak band in Raman spectra of this sample. The large shift in the peak position as compared to amino acid carbonyl -C=O mode which is expected to occur at ~ 1740 cm$^{-1}$ to higher wavenumbers reflects the existence of strong hydrogen bonding among different available functional groups in crystal lattice. The strong medium intensity –NH$_3^+$ deformation band and -C=C- ring vibration is observed in IR transmission spectra at 1633 cm$^{-1}$ (B3LYP), 1678, 1642 (HF) cm$^{-1}$ and in Raman spectra at 1630 cm$^{-1}$ (B3LYP), 1770 (HF) cm$^{-1}$. As we know that aromatic nitro compounds have strong absorption due to asymmetric stretching vibration of the NO$_2$ group at 1570-1485 cm$^{-1}$. Hydrogen bonding has a minor effect on the -NO$_2$ asymmetric stretching vibrations [41, 42]. In the present case, very strong bands of $-NO_2$ asymmetric stretching mode is observed at 1607 cm$^{-1}$ (B3LYP), 1578 cm$^{-1}$ (HF) in IR transmission spectra and at 1605 cm$^{-1}$ (B3LYP), 1580 (HF) cm$^{-1}$ in Raman spectra as weak peak components. The shifting of this mode to higher frequency confirms the realignment of the vibrations due to the inductive effect of H of Phenol. The $NH_3$ symmetric deformation mode is observed in IR transmission spectrum at 1546 cm$^{-1}$ (B3LYP) and 1535 cm$^{-1}$ (HF) as a very strong peak while in Raman spectra this medium intensity peak has two components at 1561, 1543 cm$^{-1}$ (B3LYP) and 1540, 1535 cm$^{-1}$ (HF). The N-H deformation vibrations are moved down to low wavenumber as these modes are affected by other nearby vibrations. DFT theoretical calculations showed that in aliphatic compounds with -$CH_3$ group, the series of bands are obtained pertaining to asymmetric and symmetric deformation modes in the region 1500-1400 cm$^{-1}$ [43-45]. These modes mainly from coupling of C-H and C-C stretching vibrations of deformed -$CH_3$ group. In the present investigation, -$CH_3$ asymmetric deformation mode is observed as medium intensity band at 1492 cm$^{-1}$ (B3LYP), 1465 (HF) in IR transmission spectra while at 1509, 1465 cm$^{-1}$ (B3LYP) and 1509, 1465 cm$^{-1}$ as weak bands in Raman bands are assigned as the asymmetric and symmetric deformation modes of methyl group. The presence of these bands exhibits the existence -$CH_3$ groups at different environment in crystal lattice. As it is well-known that aromatic nitro compounds have strong absorptions to asymmetric and symmetric stretching vibrations of the -$NO_2$ group at ~1450 and 1325 cm$^{-1}$, respectively. Hydrogen bonding has a minor effect on the -$NO_2$ asymmetric stretching vibrations. In the present case, -$NO_2$ asymmetric stretching weak bands have been observed at 1416, 1398 cm$^{-1}$ (B3LYP), 1437, 1400 cm$^{-1}$ (HF) in IR transmission spectra and at 1414 cm$^{-1}$ (B3LYP) and 1437

cm$^{-1}$ (HF) Raman spectra. The symmetric stretching modes of –NO$_2$ are obtained as very strong peaks at 1347, 1330, 1312 cm$^{-1}$ (B3LYP), 1338 (HF) in IR transmission spectra and at 1344, 1327 (B3LYP), 1339 (HF) Raman spectra. The presence of three peaks in B3LYP IR spectrum shows that nitro group acquire different environment or site symmetry in lattice. Their peak positions are sensitive to hydrogen bonding which may cause rearrangement through the inductive effect of H of phenol [46-48].

**In 1300 – 1000 cm$^{-1}$ region**

In this region, -CH$_3$ group deformation mode, C-O of phenolic group and C-OH, C=O, C=C groups symmetric stretching modes are observed as strong and medium intensity bands. A medium intensity peak at 1278 cm$^{-1}$ (B3LYP), 1223 cm$^{-1}$ (HF) in IR transmission spectra and in Raman spectra as weak peaks at 1310, 1275 cm$^{-1}$ (B3LYP), 1225 cm$^{-1}$ (HF), respectively are observed in these spectra. These peaks are emerged from the –CH$_3$ group deformation and -C-O phenol and -C-OH amino acid alcohol group symmetric stretching vibrations in the crystal. The medium intensity -C=O and –C=C- symmetric stretching vibrational peaks are obtained at 1165, 1140 cm$^{-1}$ (B3LYP), 1190, 1151 cm$^{-1}$ (HF) and 1096, 1061 cm$^{-1}$ (B3LYP), 1080 cm$^{-1}$ in IR transmission spectra respectively while these modes appear at 1170, 1137 cm$^{-1}$ (B3LYP), 1160 cm$^{-1}$ (HF). In Raman spectra, these -C=O and –C=C- symmetric stretching mode peaks are obtained as very weak and weak intensity peaks at 1170, 1137 cm$^{-1}$ (B3LYP), 1160 cm$^{-1}$ (HF) and 1093 cm$^{-1}$ (B3LYP), 1080 cm$^{-1}$ (HF) respectively. The two components in these stretching modes suggest that these two groups are present in different sites of crystal lattice.

**In 1000 – 400 cm$^{-1}$ region**

This region is comprised of the stretching vibrations of C-C ring, C-C skeletal, -NO$_2$ scissoring, bending, C-H bending, COO-, C-C-N deformation and –NO$_2$ bending modes. The C-C ring and skeletal stretching modes has been noticed as weak and very weak peaks at 905 cm$^{-1}$ (B3LYP), 930 cm$^{-1}$ (HF) and 818 cm$^{-1}$ (B3LYP), 792 cm$^{-1}$ (HF), respectively in IR transmission spectra. In Raman spectra these C-C ring and skeletal stretching modes are obtained as medium intensity peaks at 929, 903, 860 cm$^{-1}$ (B3LYP), 946 cm$^{-1}$ (HF) and 834, 799, 707 cm$^{-1}$ (B3LYP), 723 cm$^{-1}$ (HF) respectively. The –NO$_2$ scissoring mode appears as weak and very weak components at 766 cm$^{-1}$ (B3LYP), 741, 714 cm$^{-1}$ (HF) IR spectra and at 747, 721, 707 cm$^{-1}$ (B3LYP), 723 cm$^{-1}$ (HF) in Raman spectra. The –NO$_2$ bending, -C-H bending, -COO- deformation modes appear at 705, 698 cm$^{-1}$ (B3LYP), 696 cm$^{-1}$ (HF) in IR spectra and as very weak components at 678, 661, 644 cm$^{-1}$ (B3LYP), 698 cm$^{-1}$ (HF) in Raman spectra. Aromatic nitro compounds have a band of weak to medium intensity in the region 590–500 cm$^{-1}$ [45] due to the out-of-plane bending deformations mode of NO$_2$ group. In IR transmission spectrum, the rocking (out-of-plane bending deformation) vibration of –NO$_2$ has been observed at 575 cm$^{-1}$ (B3LYP) as medium intensity peak and in Raman spectrum this peak appears as weak intensity peak at 574 cm$^{-1}$ (B3LYP). The C-C-N in-phase deformation mode is observed as weak intensity peak at 532 cm$^{-1}$ (B3LYP), 553 cm$^{-1}$ (HF) in IR transmission spectrum and at 532 cm$^{-1}$ (B3LYP), 533 cm$^{-1}$ (HF) in Raman spectra. Carboxylate (COO$^-$) anion rocking vibration is obtained at 427 cm$^{-1}$ (B3LYP) IR transmission and Raman Spectra as a weak intensity feature.

**In 400 -10 cm$^{-1}$ region**

This is far infrared (FIR) region in which the lattice vibrations are obtained. In these, (B3LYP and HF) IR transmission and Raman spectra, the medium and weak intensity vibrational peaks are obtained. The C-C-N out-of-plane bending and skeletal vibrations are found at 385, 360 and 340 cm$^{-1}$ as medium and weak intensity components in (B3LYP), 366, 321 cm$^{-1}$ (HF) in IR transmission spectra. While in Raman spectra, these modes appear as weak and medium intensity peak components at 384, 367, 323 cm$^{-1}$ (B3LYP), 366, 321 cm$^{-1}$ (HF). The shift in out-of-phase vibrations of skeletal carbon to low wavenumber i.e. ~320 cm$^{-1}$ is attributed to the repulsion between the –NO$_2$ groups. The C-CH$_3$ in-plane bending vibration is obtained at 280 (weak), 263 (medium) cm$^{-1}$ (B3LYP), 267 cm$^{-1}$ (HF) in IR transmission spectra. In Raman spectra, this weak peak is observed at 263 cm$^{-1}$ (B3LYP). While C-CH$_3$ out-of-plane bending vibration is obtained as weak intensity band in IR transmission spectra at 203 cm$^{-1}$ (B3LYP), 196 cm$^{-1}$ (HF) and as weak intensity components in Raman spectra at 194, 150 cm$^{-1}$ (B3LYP) and 196 cm$^{-1}$ (HF). The deviation in peak position of these vibrations indicates the repulsion between methyl groups. -NO$_2$ twisting vibrations are observed at 60, 50 and 40 cm$^{-1}$. In these IR transmission spectra, these modes appear as very, very weak (vvw) components while in Raman B3LYP spectrum, 60 (vvw) and 50 (w) cm$^{-1}$ components are observed. It means one band of in-plane bending pushed up and entire bands of out of plane bending are within the region mainly due to OH group.

This in-depth IR transmission and Raman theoretical spectra from 4000-10 cm$^{-1}$ region analysis revealed the existence of *L*-threoninium and picric acid salt in ionic as well as in covalent state. These spectra showed all the peaks of functional and skeletal groups present in the crystal lattice. All the observed modes are tentatively assigned on the basis earlier reported results and available literature. The presence of hydrogen bonding cause shifting of peaks to higher or lower wavenumbers depending upon the electron inductive or withdrawing nature of the bonded functional groups present in the crystal.

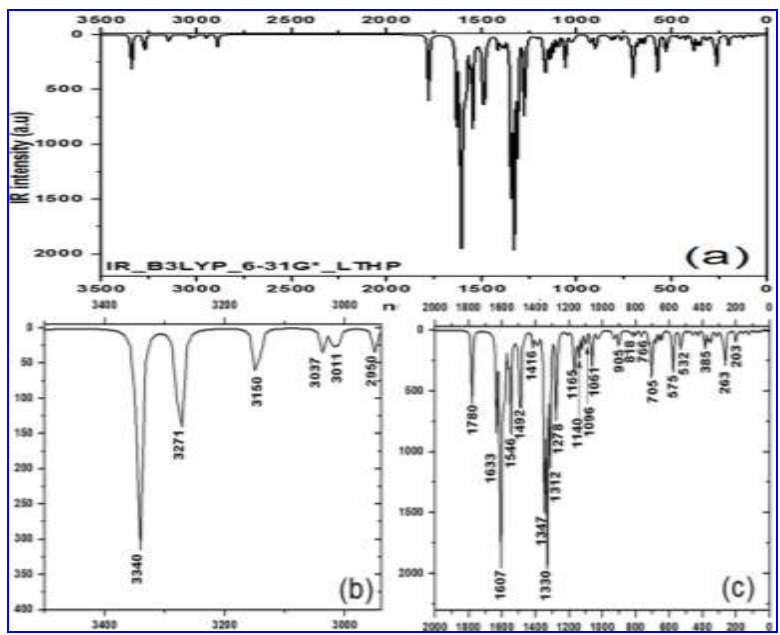

**Fig. 2**. Calculated IR spectra of L-threonine picrate (LTHP) molecule at B3LYP/6-31G* level of theory.

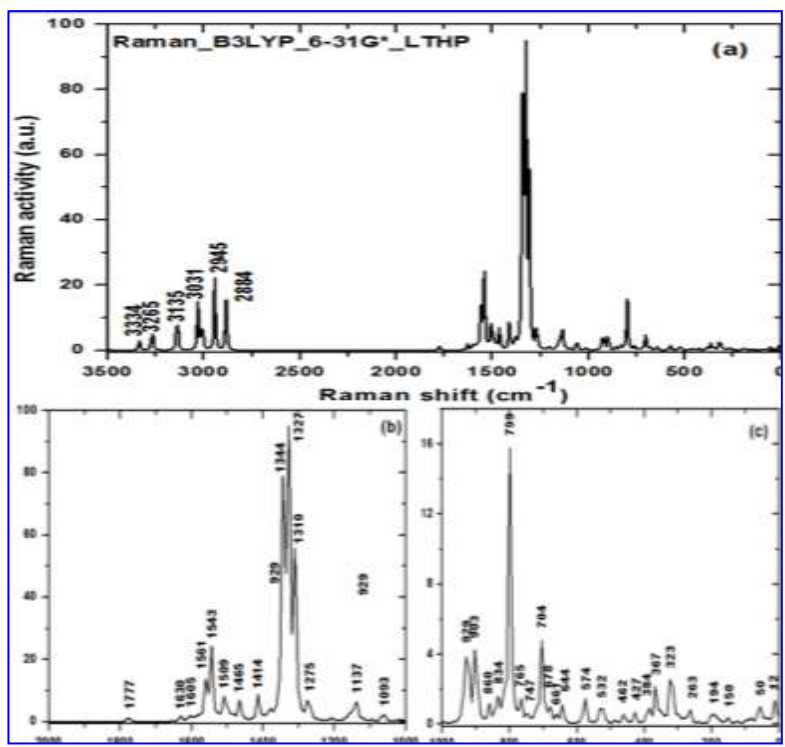

**Fig. 3**. Calculated Raman spectra of L-threonine picrate (LTHP) molecule at B3LYP/6-31G* level of theory.

## 3.3. Electro-optical properties

### 3.3.1. UV-Vis spectral analysis (TD-DFT Study)

Nowadays the time dependent range separated density functional theory is the most popular method for ground state electronic structure calculations in quantum chemistry and solid state physics [20, 21]. When compared to traditional ab initio and semi-empirical approaches the contemporary density functional methods show a favorable balance between accuracy and computational efficiency.

To analyze the nature of electronic transition in LTHP molecule the TD-DFT TD-B3LYP/6-31G$^*$ and range separated functionals CAM/B3LYP/6-31G$^*$ and LC-BLYP/6-31G$^*$ levels of theory. Fig. 4 shows the calculated UV-Vis spectra of LTHP molecule using different functionals such as B3LYP, LC-BLYP and CAM-B3LYP using 6-31G* basis set. Fig. 4 shows two absorption maxima at 210 and 281 nm calculated at TD-B3LYP/6-31G* level of theory. However, these values dramatically changed when we used range separated functionals such as LC-BLYP and CAM-B3LYP which are found to be 237, 323 nm and 262, 351 nm, respectively. The calculated values from CAM-B3LYP are found to be very close to the experimental results [16]. Optimized geometry of LTHP molecule represents that the maximum absorption wavelength is due to the electronic transition from HOMO to LUMO which will be explained in the next section. The observed bands are due to $\pi - \pi^*$ and n– $\pi^*$ transitions and in the chemical stability of LTHP molecule both HOMO and LUMO are taking participation.

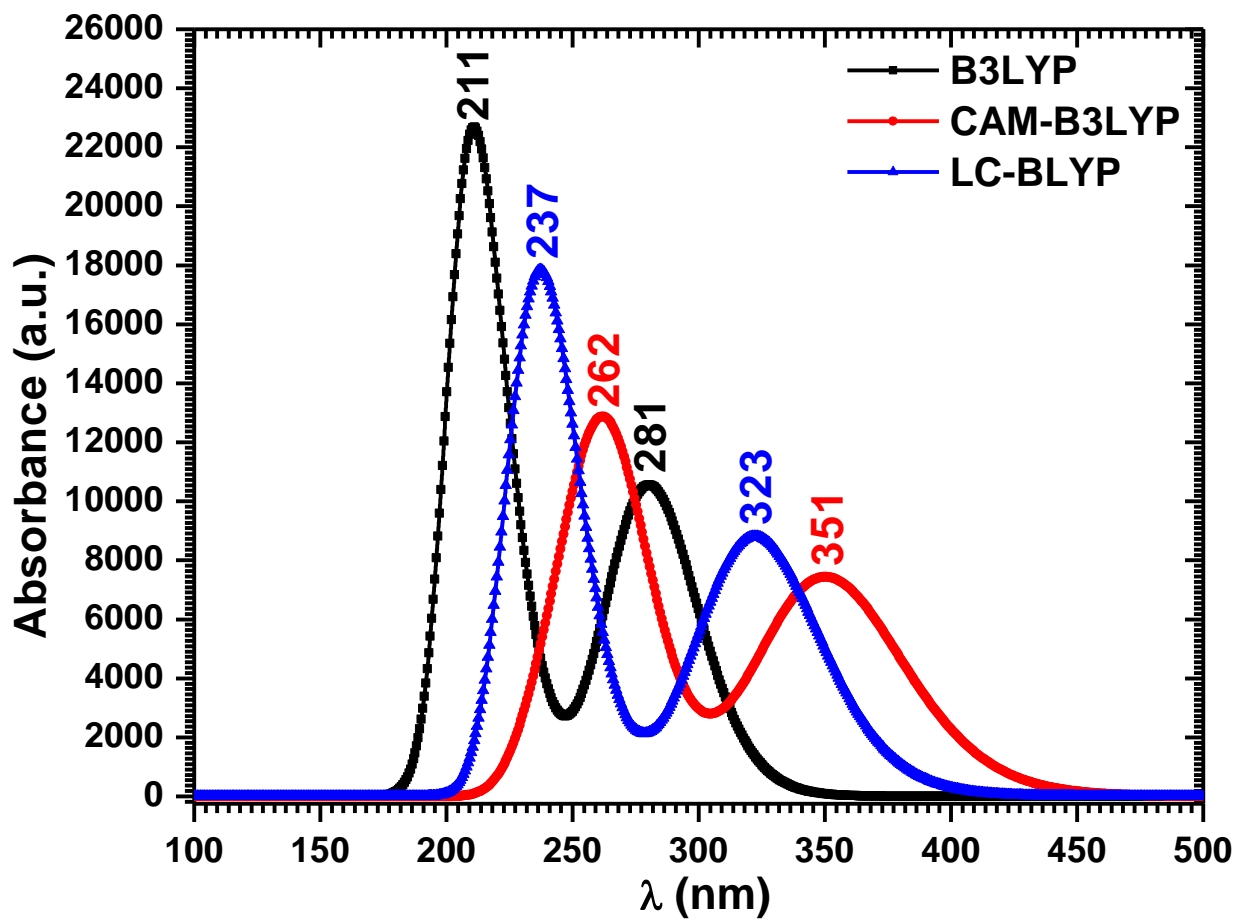

**Fig 4**. Calculated UV-Visible spectra of LTHP molecule at different levels of theory.

*3.3.2. Electronic properties*

The frontier molecular orbitals (FMOs) such as HOMO and LUMO, represents the ability to donate or accept the electrons, respectively in any compound which plays important role in its electrical and optical properties as well as chemical reactions. The chemical hardness, potential, reactivity, softness, electronegativity, electrophilicity, kinetic stability and optical Polarizability, etc. of any molecule can be explained with the help of HOMO and LUMO energies. The energy values of HOMO and LUMO orbitals calculated by B3LYP using 6-31G* basis set are presented in Table S3 (see supplementary data).

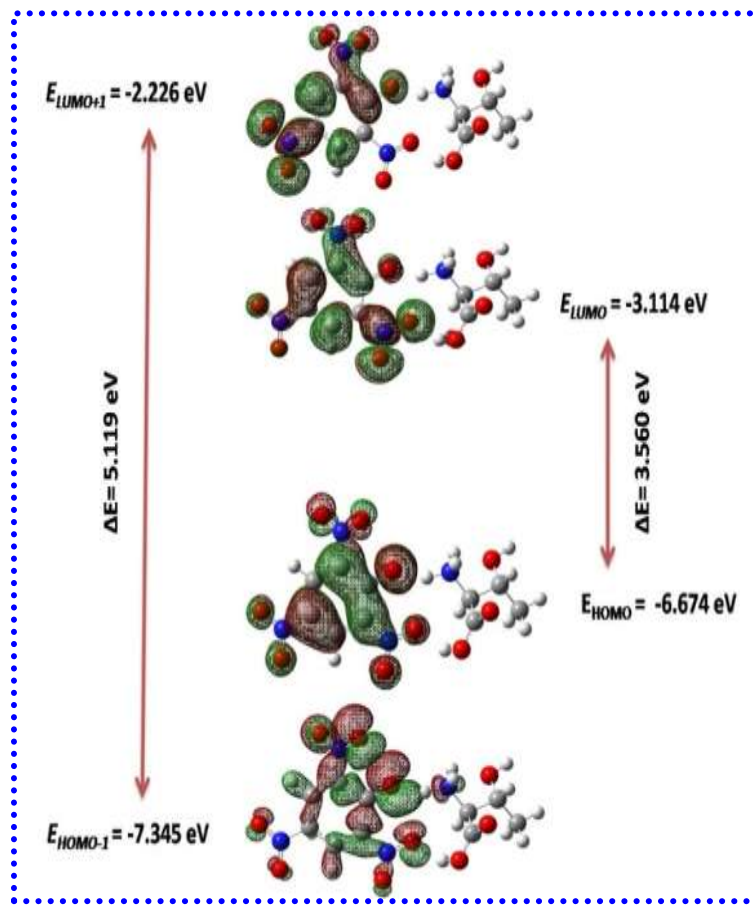

**Fig 5**. The HOMO, HOMO-1, LUMO and LUMO+1 representation and charge density distribution of LTHP molecule calculated at B3LYP/6-31G* level of theory.

The atomic orbital composition of the FMOs are sketched and shown in Fig. 5. From the Fig. 5, it is clear that the HOMO and LUMO both are distributed on picrate anion. From the Mullikan charge analysis it is also clear that the charge transfer take place mainly within the picrate anion upon excitation. The density of states (DOS) play vital role to understand the energy spectrum especially the electron distribution. To shed light on the electronic band structure of LTHP, we also studied the total/partial DOS (T/PDOS) using GGA/BLYP and presented in Fig. 6. In LTHP, the energy band in the bottom of valence band (VB) at ~ -18.5 to -19.0, -14.0 Hartree, the middle of VB at ~ -9.5 to -10.0 Hartree is defined by the *s*-orbitals. Moreover, the *s*-orbitals further appear in the upper VB and conduction band (CB) in the energy range of 0 to -1.0 and 0 to 0.4 Hartree, respectively, which reveals the negligible contribution of *s*-orbitals into the electronic properties of DPNDF. The upper VB and CB in the energy range ~ 0 to -1.0 and 0 to 0.4 are only contributions from the *p*-orbitals. The DOS profiles of the LTHP shows the upper VB and CB are primarily of *s* and *p*-character.

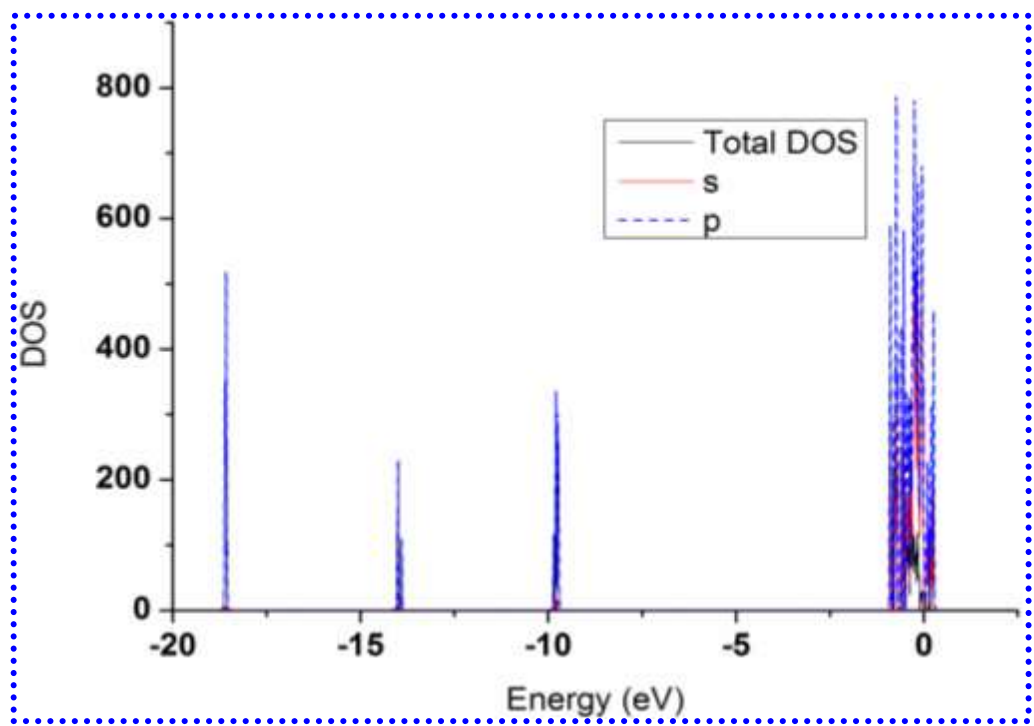

**Fig. 6.** Density of states of LTHP molecule.

The molecular electrostatic potential (MEP) mapping of LTHP has been illustrated in Fig. 7. The MEP is helpful property to investigate the reactivity of molecular species by predicting the possible site for nucleophile and electrophilic attack. In MEP plot, while maximum positive region that is preferred site for nucleophilic attack indicated as blue color. Similarly, a maximum negative region is preferred site for electrophilic attack that is indicated as red surface. The TED plots of the titled molecule show a uniform distribution.

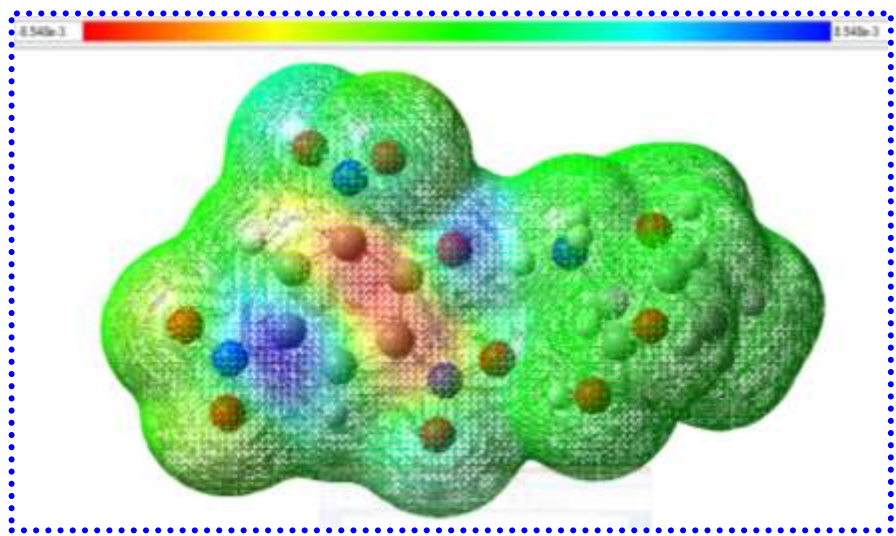

**Fig. 7**. Molecular electrostatic potential (MEP) map of LTHP molecule at ISO value 0.02.

Global chemical reactivity descriptors *(GCRD)* are based on the conceptual DFT and are used to understand the relationship between global chemical reactivity, structure, stability etc. parameters of any molecule. Additionally, these descriptors are also employed in the progress of quantitative structure-activity (QSAR), quantitative structure-property (QSPR), and quantitative structure-toxicity (QSTR) relationships. The hardness of any molecule is related to aromaticity [49]. The DFT provides the definitions of important universal concepts of molecular structure stability and reactivity [50]. Here our focus is to find the values of various GCRD parameters such as chemical hardness ($\eta$), chemical potential ($\mu$), softness (S), electronegativity ($\chi$) and electrophilicity index ($\omega$) of LTHP molecule using HOMO and LUMO energies values using different functionals which are given in Table S3 (see supplementary data). The calculated HOMO-LUMO energy gap and chemical hardness values show that the titled molecule possesses good chemical stability. The larger energy gap suggests that the LTHP has significant kinetic stability as according to softness-hardness rule when the molecule possesses wide energy gap then said to be hard and also show higher thermal and kinetic stability.

## 3.4. Polarizability and first hyperpolarizability studies

Nonlinear activity of any materials provides the key functions from applications point of views for frequency shifting, optical modulation, optical switching etc. for developing technologies in the area of communication, optical processing and interconnection [51, 52]. It is well known that the importance of polarizability and hyperpolarizabity of a molecular system is dependent on the electronic communication of two different parts of a molecule. There are many reports available on the choice of suitability of functional for nonlinear optical parameter calculations [20, 21, 53, 54]. We have chosen the various functionals like: B3LYP and range separated CAM-B3LYP, LC-BLYP using 6-31G$^*$ basis set and various parameters such as electronic dipole moment ($\mu$), total average polarizability ($\alpha_0$), anisotropy of polarizability ($\Delta\alpha$) and static first hyperpolarizability ($\beta_{tot}$, $\beta_0$) have been calculated and presented in Table 2 (computational details can be found in supporting information) and also the variation of $\mu_{tot}$ and $\beta_{tot}$ calculated by above mentioned methods with respect to urea is shown in Figure 2S(a) and (b) [see supporting information].

The value of total dipole moment ($\mu_{tot}$) and first hyperpolarizability is found to be 3 and 51 times higher than urea ($\mu_{Urea}$= 4.24 Debye, $\beta_{tot}$ = 0.22 ×10$^{-30}$ esu.) molecule calculated at B3LYP/6-31G* level of theory [55] as well as many other compounds [56-63], while experimentally the second harmonic generation of the titled compound in solid form was measured by S. Natarajan et al., and found to be 43 times higher than KDP molecule [16]. Such high nonlinearity recommended the titled compound for optoelectronic and electro-optic device applications.

**Table 2** Calculated values of polarizabilities, dipole moments and hyperpolarizability along with their individual tensor components by DFT/B3LYP, CAM-B3LYP and LC-BLYP using /6-31G* basis set of LTHP molecule.

| | Polarizabilities and dipole moments | | | | | | | Hyperpolarizabilities | | | | | |
|---|---|---|---|---|---|---|---|---|---|---|---|---|---|
| | B3LYP | | CAM-B3LYP | | LC-BLYP | | | B3LYP | | CAM-B3LYP | | LC-BLYP | |
| | a.u. | esu ($\times 10^{-24}$) | a.u. | esu. ($\times 10^{-24}$) | a.u. | esu ($\times 10^{-24}$) | | a.u. | esu ($\times 10^{-30}$) | a.u. | esu. ($\times 10^{-30}$) | a.u. | esu. ($\times 10^{-30}$) |
| $\alpha_{xx}$ | 264.989 | 39.271 | 249.916 | 37.038 | 239.668 | 35.519 | $\beta_{xxx}$ | -616.094 | -5.316 | -469.490 | -4.051 | -395.264 | -3.411 |
| $\alpha_{xy}$ | 2.697 | 0.399 | 2.837 | 0.421 | 3.2705 | 0.485 | $\beta_{xxy}$ | -789.778 | -6.815 | -729.921 | -6.298 | -707.328 | -6.104 |
| $\alpha_{yy}$ | 200.641 | 29.735 | 192.177 | 28.481 | 186.106 | 27.581 | $\beta_{xyy}$ | 131.381 | 1.134 | 167.528 | 1.446 | 213.629 | 1.844 |
| $\alpha_{xz}$ | 9.811 | 1.454 | 9.397 | 1.393 | 9.522 | 1.411 | $\beta_{yyy}$ | 345.620 | 2.982 | 271.682 | 2.344 | 206.938 | 1.786 |
| $\alpha_{yz}$ | 5.648 | 0.837 | 5.343 | 0.792 | 5.072 | 0.752 | $\beta_{xxz}$ | -111.222 | -0.960 | -91.173 | -0.787 | -81.141 | -0.7 |
| $\alpha_{zz}$ | 97.191 | 14.404 | 95.708 | 14.184 | 94.532 | 14.01 | $\beta_{xyz}$ | -68.439 | -0.591 | -57.278 | -0.495 | -51.170 | -0.442 |
| $\alpha_{tot}$ | 187.607 | 27.803 | 179.267 | 26.567 | 173.435 | 25.703 | $\beta_{yyz}$ | 48.124 | 0.415 | 38.025 | 0.328 | 35.725 | 0.308 |
| $\Delta\alpha$ | 147.608 | 21.876 | 135.922 | 20.144 | 128.186 | 18.997 | $\beta_{xzz}$ | -5.413 | -0.047 | -5.871 | -0.051 | -5.149 | -0.045 |
| $\mu_x$ | 4.605 | 11.697D | 4.530 | 11.514D | 4.535 | 11.528D | $\beta_{yzz}$ | 7.095 | 0.061 | 1.713 | 0.015 | 0.660 | 0.006 |
| $\mu_y$ | 0.781 | 1.980D | 0.692 | 1.758D | 0.651 | 1.656D | $\beta_{zzz}$ | -22.020 | -0.190 | -16.955 | -0.146 | -14.665 | -0.127 |
| $\mu_z$ | -1.116 | -2.835D | -1.142 | -2.903D | -1.175 | -2.988D | $\beta_0$ | 783.353 | 6.760 | 632.898 | 5.642 | 542.233 | 4.679 |
| $\mu_{tot}$ | 4.80 | 12.20D | 4.723 | 12.003D | 4.729 | 12.023D | $\beta_{tot}$ | 1305.588 | 11.266 | 1054.981 | 9.103 | 903.721 | 7.798 |

Fo $\alpha$, 1 a.u. = 0.1482×10$^{-24}$ esu, for $\beta$, 1 a.u. = 0.008629×10$^{-30}$ esu, $\mu_{Urea}$= 4.24D and $\beta_{Urea}$ = 0.22 ×10$^{-30}$ esu [55]

## 4. Conclusions

The strong hydrogen bonding was observed between picric acid and L-threonine. The proton transfer from picric acid to L-threonine is confirming the formation of L-threoninium picrate compound. The HF and B3LYP calculated infrared and Raman frequencies were assigned and compared with the experimental values and found to be in close approximation. The dipole moment ($\mu_{tot}$), polarizability ($\alpha_0$) and first hyperpolarizability ($\beta_{tot}$) values were calculated at different levels of theory. The $\beta_{tot}$ value was found to be 51 times higher than urea computed at same level of theory. Two absorption bands were observed in UV-Vis spectra at all the applied methods. The absorption bands calculated using range separated functional CAM-B3LYP were found at 262 and 351 nm. The total/partial DOS (T/PDOS) was determined using GGA/BLYP. The chemical reactivity descriptors show that LTHP has significant chemical stability. The molecular electrostatic potential map gives the information about nucleophilic attack electrophilic attack in the molecule. Such high nonlinear property of LTHP molecule suggest that it is a good candidate for higher order nonlinear applications and also may help to the scientists and researchers in design and synthesis of new materials with unique optical properties as well.

*Acknowledgment*

*The authors are thankful to King Khalid University, Abha, Saudi Arabia for providing research infrastructure. The authors are also thankful to Dr. I.S. Yahia, Nano-Science & Semiconductor Labs., Faculty of Education, Ain Shams University, Roxy, Cairo, Egypt, for supporting the calculations with Gaussian software.*

**Molecular structure, vibrational, photophysical and nonlinear optical properties of L-threoninium picrate: A first-principles study**

*S. AlFaify[a]\*, Mohd. Shkir[a]\*, Manju Arora[b], Ahmad Irfan[c,d], H. Algarni[a], Haider Abbas[e], Abdullah G. Al-Sehemi[c,d]*

[a]Department of Physics, Faculty of Science, King Khalid University, P.O. Box. 9004 Abha 61413, Saudi Arabia

[b]CSIR- National Physical Laboratory, Dr. K.S. Krishnan Road, New Delhi 110012, India

[c]Department of Chemistry, Faculty of Science, King Khalid University, Abha 61413, P.O. Box 9004, Saudi Arabia

[d]Research Center for Advanced Materials Science, King Khalid University, Abha 61413, P.O. Box 9004, Saudi Arabia

[e]Department of Physics, Manav Rachna College of Engineering, Faridabad, Haryana 121001, India

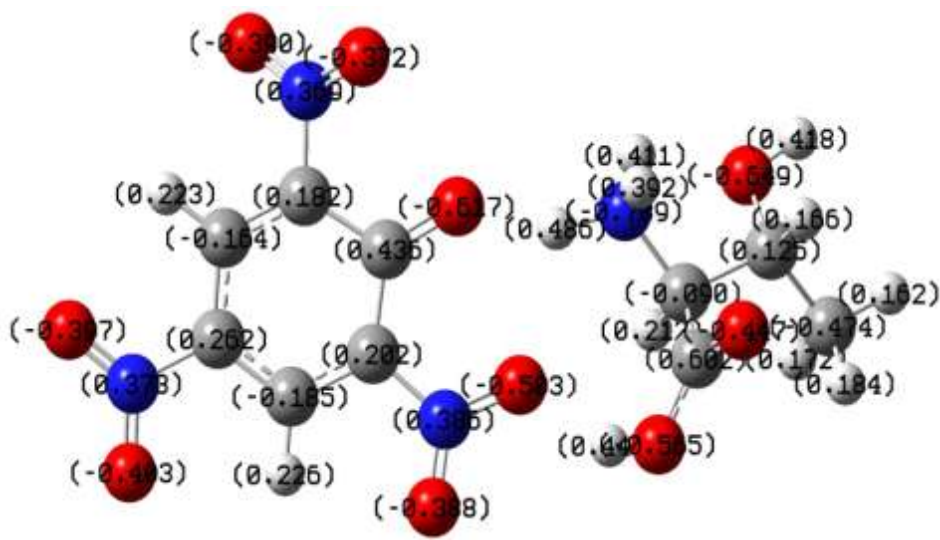

Fig. S1. Mulliken charges of LTHP molecule

**Table S1** Mulliken charge analysis of L-threonine picrate (LTHP) molecule.

| Atom No. | Atom | Mulliken charge |
|---|---|---|
| 1 | C | 0.60181 |
| 2 | C | -0.09022 |
| 3 | H | 0.21167 |
| 4 | C | 0.12460 |
| 5 | H | 0.16569 |
| 6 | C | -0.47418 |
| 7 | H | 0.17241 |
| 8 | H | 0.18383 |
| 9 | H | 0.16228 |
| 10 | N | -0.78857 |
| 11 | H | 0.48554 |
| 12 | H | 0.41110 |
| 13 | H | 0.39242 |
| 14 | O | -0.44743 |
| 15 | O | -0.56457 |
| 16 | O | -0.64856 |
| 17 | H | 0.41760 |
| 18 | C | 0.43593 |
| 19 | C | 0.18196 |
| 20 | C | -0.16405 |
| 21 | H | 0.22276 |
| 22 | C | 0.26184 |
| 23 | C | -0.18510 |
| 24 | H | 0.22591 |
| 25 | C | 0.20222 |
| 26 | N | 0.36907 |
| 27 | N | 0.37302 |
| 28 | N | 0.38573 |
| 29 | O | -0.61733 |
| 30 | O | -0.37246 |
| 31 | O | -0.38985 |
| 32 | O | -0.39731 |
| 33 | O | -0.40261 |
| 34 | O | -0.38779 |
| 35 | O | -0.50334 |
| 36 | H | 0.44600 |

**Table S2** Calculated IR and Raman frequencies with their assignments of LTHP obtained by DFT using B3LYP/6-31G* and HF/6-31G* levels of theory.

| Reported [22] | | Current work | | | |
|---|---|---|---|---|---|
| | | B3LYP | | HF | |
| IR frequencies (cm$^{-1}$) | Assignments | IR frequencies (cm$^{-1}$) | Raman Frequencies (cm$^{-1}$) | IR frequencies (cm$^{-1}$) | Raman Frequencies (cm$^{-1}$) |
| 3400 | O-H stretching | 3340 | 3334 | 3348 | 3348 |
| 3238 | N-H, NH$_2$ symmetric stretching | 3271, 3150 | 3265,3135 | 3276 | 3277 |
| 2914 | C-H stretching | 3037,3011,2920 | 3031,2945,2884 | 3089,2837 | 3090,2980 |
| 1745 | C=O Asymmetric stretching | 1780 | 1777 | 1810 | 1812 |
| 1612 | NH$_3^+$ Asymmetric deformation, C=C ring vibration | 1633 | 1630 | 1678,1642 | 1770 |
| 1570 | NO$_2$ Asymmetric stretching | 1607 | 1605 | 1578 | 1580 |
| 1541 | NH$_3$ symmetric stretching | 1546 | 1561,1543 | 1535 | 1540,1535 |
| 1493 | CH$_3$ Asymmetric deformation | 1492 | 1509,1465 | 1465 | 1509,1465 |
| 1425 | NO$_2$ stretching | 1416,1398 | 1414 | 1437,1400 | 1437 |
| ----- | NO$_2$ symmetric stretching | 1347,1330,1312 | 1344,1327 | 1338 | 1339 |
| 1273 | CH$_3$ deformation, C-O phenolic, symmetric (C-OH) stretching | 1278 | 1310, 1275 | 1223 | 1225 |
| 1163 | C=O stretching | 1165,1140 | 1170,1137 | 1190,1151 | 1160 |
| 1098 | C=C stretching | 1096,1061 | 1093 | 1080 | 1080 |
| 918 | C-C stretching/ring | 905 | 929,903,860 | 930 | 946 |
| 787 | C-C skeletal stretching | 818 | 834,799,765 | 792 | 805 |
| 743 | NO$_2$ scissoring | 766 | 747,721,707 | 741,714 | 723 |
| 706 | NO$_2$ bending, C-H bending, COO$^-$ deformation | 705,698 | 678,661,644 | 696 | 698 |
| 542 | C-C-N deformation, ρ(NO$_2$) | 532 | 532 | 553 | 535 |
| — | Skeletal deformation/lattice vibrations | 385,263,203 | 427,367,323,263, 194,150,55,12 | 450, 366,267, 196 | 450,366,321, 196 |

**NLO properties methodology**

For calculating the above said parameters the mathematical expressions are defined as:

Dipole moment

$$\mu_{tot} = (\mu_x^2 + \mu_y^2 + \mu_z^2)^{\frac{1}{2}} \tag{3}$$

Total average polarizability

$$\alpha_0 = \frac{1}{3}(\alpha_{xx} + \alpha_{yy} + \alpha_{zz}) \tag{4}$$

Anisotropy of polarizability

$$\Delta\alpha = \frac{1}{\sqrt{2}}\sqrt{\left[(\alpha_{xx} - \alpha_{yy})^2 + (\alpha_{yy} - \alpha_{zz})^2 + (\alpha_{zz} - \alpha_{xx})^2 + 6\alpha_{xz}^2\right]} \tag{5}$$

Components of first hyperpolarizability can be calculated by using the expression

$$\beta_i = \beta_{iii} + \sum_{i \neq j}\left[\frac{(\beta_{ijj} + 2\beta_{jii})}{3}\right] \tag{6}$$

Using x, y, z components the magnitude of first hyperpolarizability ($\beta_{tot}$) can be calculated as:

$$\beta_{tot} = \sqrt{(\beta_x^2 + \beta_y^2 + \beta_z^2)} \tag{7}$$

*Where $\beta_x$, $\beta_y$ and $\beta_z$ are:*

$\beta_x = (\beta_{xxx} + \beta_{xxy} + \beta_{xyy})$

$\beta_y = (\beta_{yyy} + \beta_{xxz} + \beta_{yyz})$

$\beta_z = (\beta_{xzz} + \beta_{yzz} + \beta_{zzz})$

Hence, the final equation for magnitude of total first static hyperpolarizability calculation is given by

$$\beta_{tot} = \sqrt{\left[(\beta_{xxx} + \beta_{xxy} + \beta_{xyy})^2 + (\beta_{yyy} + \beta_{xxz} + \beta_{yyz})^2 + (\beta_{xzz} + \beta_{yzz} + \beta_{zzz})^2\right]} \tag{8}$$

The first hyperpolarizability can be described by a 3 × 3 × 3 matrix as it is known as a third order tensor and 27 components of the 3D matrix are reduced to 10 components due to the Kleinman symmetry [64], and can be given in lower tetrahedral format. It is obvious that the lower part of 3 × 3 × 3 matrices is tetrahedral. The static first hyperpolarizability was also calculated using the following relation, $\beta_0 = \frac{3}{5}\beta_{tot}$. The calculated values of 6 components of polarizability, 3 components of dipole moment and 10 components of hyperpolarizability along with their total values with or without conversion i.e. a.u. and esu., are presented in Table 2, as for α, 1 a.u.= $0.1482\times10^{-24}$ esu and for β, 1 a.u.=$0.008639\times10^{-30}$. The electronic communication of two different parts of any compound plays a vital role in polarizability and hyperpolarizabity values. The variation of total dipole moment and first hyperpolarizability values calculated by B3LYP, CAM-B3LYP and LC-BLYP functional in comparison of reported value of urea are shown in Fig. S2 (a) and (b).

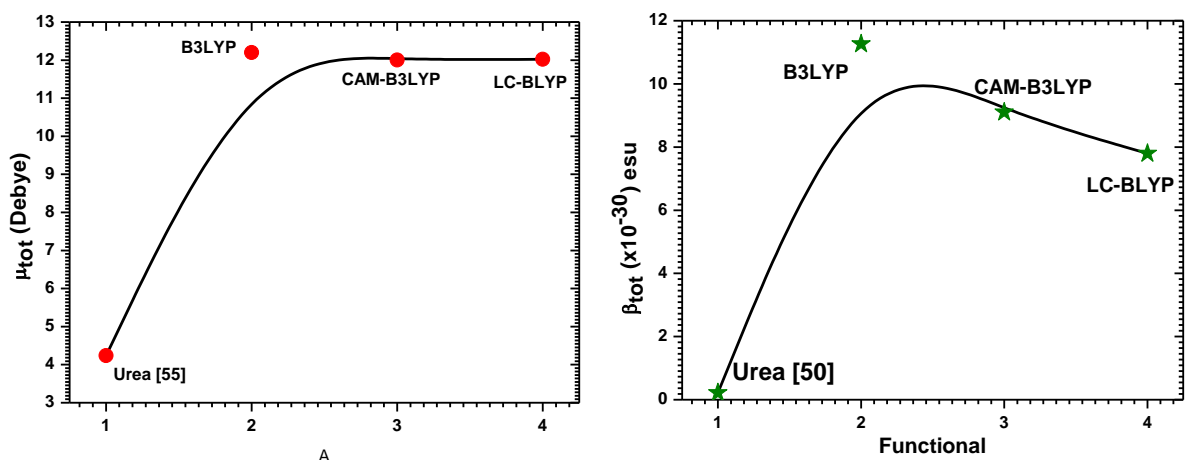

Fig. S2. Variation of (a) dipole moment and (b) first hyperpolarizability calculated by different functionals at 6-31G* basis set.

**Global chemical reactivity descriptors**

An approximation for absolute hardness η was developed [65-67] as given below:

$$\eta = \frac{I-A}{2} \qquad (8)$$

where I is the vertical ionization energy and A is vertical electron affinity.

As per Koopmans theorem [68] the ionization energy and electron affinity can be specified through HOMO and LUMO orbital energies as:

I = −E$_{HOMO}$

A = −E$_{LUMO}$

Values of I and A of LTHP molecules calculated at both level of theories are given in Table S3. The higher energy of HOMO is corresponds to the more reactive molecule in the reactions with electrophiles, while lower LUMO energy is essential for molecular reactions with nucleophiles [69].

Hence, the hardness of any materials is corresponds to the gap between the HOMO and LUMO orbitals. If the energy gap of HOMO-LUMO is larger then molecule would be harder [66].

$$\eta = \frac{1}{2}(E_{LUMO} - E_{HOMO}) \qquad (9)$$

The electronic chemical potential of a molecule is calculated by:

$$\mu = -\left(\frac{I+A}{2}\right) \qquad (10)$$

The softness of a molecule is calculated by:

$$S = \frac{I}{2\eta} \qquad (11)$$

The electronegativity of the molecule is calculated by:

$$\chi = \left(\frac{I+A}{2}\right) \qquad (12)$$

The electrophilicity index of the molecule is calculated by:

$$\omega = \frac{\mu^2}{2\eta} \qquad (13)$$

The calculated values of GCRD such as η, µ, S, χ and ω for LTHP molecule are also presented in Table S3.

**Table S3**

Calculated energy values of FMOs, their difference, *η, µ, S, χ* and *ω* at B3LYP/6-31G* basis set for LTHP molecule in G.S.

| Parameters | B3LYP ( e.V.) |
|---|---|
| $E_{HOMO}$ | -6.674 |
| $E_{HOMO-1}$ | -7.345 |
| $E_{LUMO}$ | -3.114 |
| $E_{LUMO+1}$ | -2.226 |
| $\Delta E_{HOMO-LUMO}$ | 3.559 |
| $\Delta E_{HOMO-1-LUMO+1}$ | 5.119 |
| Hardness (η) | 1.780 |
| Potential (µ) | -4.894 |
| Softness (S) | 1.875 |
| Electronegativity (χ) | 4.894 |
| Electrophilic index (ω) | 6.728 |